\documentclass[iop]{emulateapj}
\usepackage{graphicx}
\usepackage{subfigure}
\usepackage{hyperref}
\usepackage{epsfig}
\usepackage{natbib}
\usepackage{multirow}
\usepackage{amsmath}
\usepackage{courier}

\providecommand{\e}[1]{\ensuremath{\times 10^{#1}}} 

\begin{document}

\title{Constraining White Dwarf Structure and Neutrino Physics in 47 Tucanae}
\author{ R.~Goldsbury\altaffilmark{1}, J.~Heyl\altaffilmark{1}, H.~B.~Richer\altaffilmark{1}, J.~S.~Kalirai\altaffilmark{2,3}, P.E. Tremblay\altaffilmark{4}}

\altaffiltext{1}{Department of Physics \& Astronomy, University of British Columbia, Vancouver, BC, Canada V6T 1Z1}
\altaffiltext{2}{Space Telescope Science Institute, 3700 San Martin Drive, Baltimore, MD, 21218}
\altaffiltext{3}{Center for Astrophysical Sciences, Johns Hopkins University, Baltimore, MD, 21218}
\altaffiltext{4}{Department of Physics, University of Warwick, Coventry, CV4 7AL, UK}

\begin{abstract}

We present a robust statistical analysis of the white dwarf cooling sequence in 47 Tucanae. We combine HST UV and optical data in the core of the cluster, Modules for Experiments in Stellar Evolution (MESA) white dwarf cooling models, white dwarf atmosphere models, artificial star tests, and a Markov Chain Monte Carlo (MCMC) sampling method to fit white dwarf cooling models to our data directly. We use a technique known as the unbinned maximum likelihood to fit these models to our data without binning. We use these data to constrain neutrino production and the thickness of the hydrogen layer in these white dwarfs. The data prefer thicker hydrogen layers $(q_\mathrm{H}=3.2\e{-5})$ and we can strongly rule out thin layers $(q_\mathrm{H}=10^{-6})$. The neutrino rates currently in the models are consistent with the data. This analysis does not provide a constraint on the number of neutrino species.

\end{abstract} 

\section{Introduction}

White dwarfs are often utilized as a way to indirectly test physical properties that are not directly observable themselves. White dwarfs have been used to measure the distances \citep{renzini1996white,zoccali2001white,woodley2012spectral}, and ages \citep{hansen2007white,hansen2013age} of globular clusters. They have been used to constrain the initial mass function and age of the galaxy \citep{wood1992constraints}. The physics of crystallization of dense white dwarf interiors has been indirectly constrained by \cite{winget2009physics}. The white dwarf cooling sequence has even been used to measure dynamical relaxation in 47 Tucanae \citep{heyl2015measurement}. White dwarfs in globular clusters have proven particularly useful for this, because all of the objects in a cluster come from an initial population with roughly the same age, extinction, distance, and typically the same initial chemical composition.

In this paper we use the most well-populated white dwarf sequence ever measured in a globular cluster (47 Tuc) to indirectly constrain the thickness of the Hydrogen layer, and the rate of neutrino production in the cores of white dwarfs. This will build on our previous work from \cite{goldsbury2012empirical}, in which we constrained white dwarf cooling models from various groups against multi-band HST photometry in 47 Tuc. These models are again compared to the new data set in Section \ref{othergroups}.

Our dataset is obtained from nine HST orbits that cover the core of 47 Tuc in ultraviolet and visible filters. In total more than 90\% of the inner 5 arc-minutes of the cluster are observed, and over 3500 white dwarfs are detected. This dataset is described in detail in Section \ref{obs}. The analysis discussed in this paper complements the work in \cite{hansen2015constraining}. In that paper, similar white dwarf cooling models were fit to the ACS data shown as the surrounding annulus in Figure \ref{obs_pattern}. These models were fit by binning the data in two dimensions (color and magnitude) and using the $\chi^2$ statistic to address the quality of the fit.

Our analysis combines data in separate regions from both WFC3 and ACS images. To take full advantage of the constraining power of this dataset, we use an unbinned maximum likelihood method. The data occupy a three dimensional parameter space comprised of magnitudes in two filters and a radial coordinate in the observed field. Theoretical models are transformed into this space using the measured properties of the detection instrument and the reduction procedures. The data are modified as little as possible and our assumptions about detection efficiency and photometric error are built directly into the model itself. This process is described in detail in Section \ref{ana}.

\section{Data}
\label{obs}

\begin{figure}[htbp]
\centering
\epsfig{file=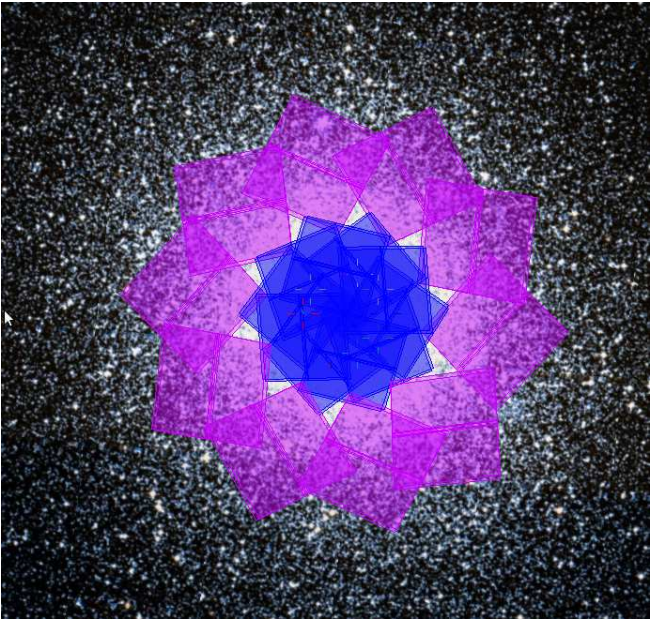,width=\linewidth}
\caption{Above are the pattern of the observed fields in the core of 47 Tuc. Fields imaged by WFC3 are shown in blue. Those observed by ACS are in pink. In the final dataset only nine of the ten orientations for both WFC3 and ACS were used due to loss of guide stars during one exposure.}
\label{obs_pattern}
\end{figure}

Our dataset results from ten total HST orbits during Cycle 21 (GO-12971, PI: Richer), one of which was rejected due to loss of guide stars. Observations were taken between November 2012 and September 2013. In each orbit, the cluster was simultaneously observed using the Wide Field Camera 3 (WFC3) and the Advanced Camera for Surveys (ACS). All of the observed fields are shown in Figure \ref{obs_pattern}. Each orbit was split into two exposures for each of the four filters. These are listed in Table \ref{exptable}. Splitting each observation in two allows us to eliminate cosmic ray strikes. The exposure lengths are chosen to fit nicely around the required buffer memory dumps from the telescope. The observations are designed to take advantage of the maximum possible observation time during each orbit. The split observation also allows for a dither between each exposure.

\begin{table}
\caption{Exposure times in seconds, for each of the nine visits}

\centering
\begin{tabular}{| c | c || c | c |}

\hline
Camera & Filter & Exposure 1 & Exposure 2\\
\hline

WFC3 & F225W & 380 & 700\\
 	 & F336W & 720 & 485\\
 \hline
ACS  & F435W & 290 & 690\\
     & F555W & 660 & 360\\

\hline
\end{tabular}
\label{exptable}
\end{table}

\begin{figure*}[h!tbp]
\centering
\epsfig{file=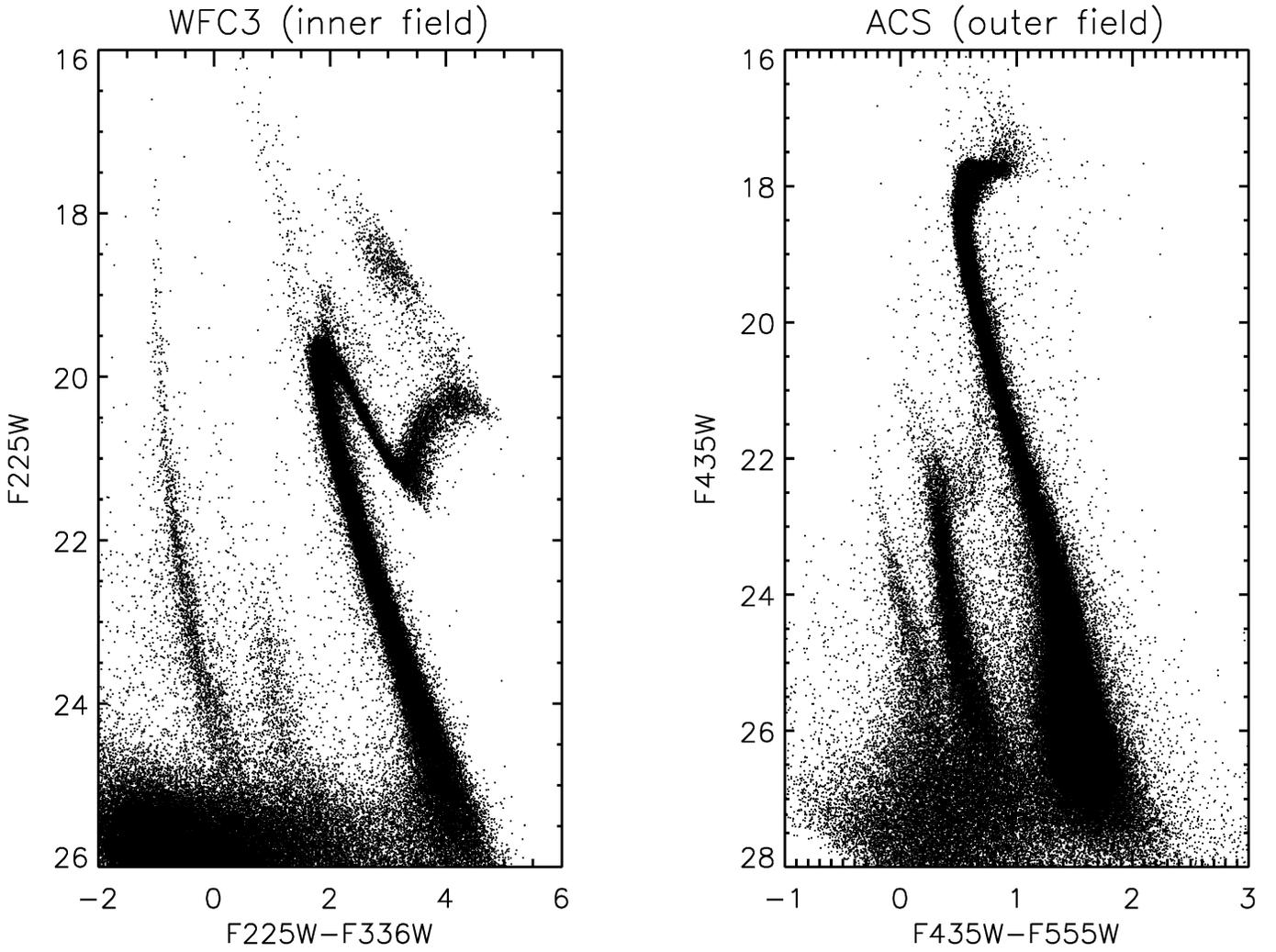,width=\linewidth}
\caption{The color magnitude diagrams of the two fields. The central UV data are shown on the left. The visible CMD from the surrounding annulus is shown on the right.}
\label{dualcmd}
\end{figure*}

The raw WFC3 images were combined by drizzling onto a single reference frame \citep{fruchter2009multidrizzle}. The ACS images were reduced to photometry independently. In total 107,000 objects were detected in the WFC3 fields, and 228,000 objects were detected in the ACS fields. The CMDs resulting from these catalogs are shown in Figure \ref{dualcmd}.

\section{Analysis}
\label{ana}

\subsection{Artificial Star Tests}
\label{arts}

The process by which we generate and measure artificial stars is one of many steps that are integral to the translation of the white dwarf cooling model from theory-space into data-space. By ``theory-space" we mean the quantities that come out of the theoretical white dwarf calculations, such as their physical size, their atmospheric composition, their surface temperatures, etc. By ``data-space" we mean the dimensions of the space in which the data points exist. The WFC3 data that we are considering are the F225W magnitude, the F336W magnitude, and the radial distance in pixels from the cluster centre of every white dwarf in the sample. So the points in the data set all lie in this three-dimensional space. Similarly, the ACS data-space is made up of F435W, F555W, and radial distance. Any comparison of model to data necessarily requires the model and the data to be transformed to common quantities that are then compared. In our previous paper on white dwarf cooling \citep{goldsbury2012empirical} we took an approach that involved modifying the data to produce a time-temperature relation that was then compared to the model. In this paper we do the opposite, which is to leave the data as untouched as possible and perform all of the transformations on the model itself.

The artificial star tests for the WFC3 drizzled image are described below. The process is virtually identical for the ACS data, but is only performed on one of the nine visits. We make the assumption that the cluster's projection is azimuthally symmetric, so the error and completeness measured across one visit are the same for the eight other ACS visits.

\begin{figure*}[h!tbp]
\centering
\epsfig{file=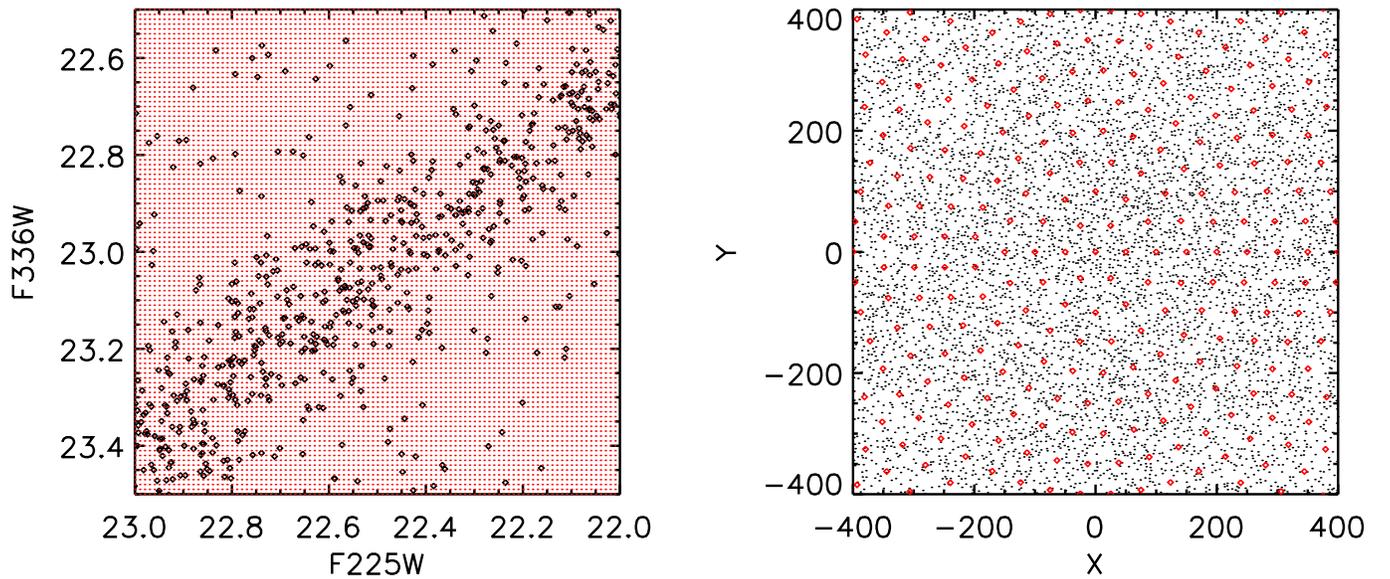,width=\linewidth}
\caption{The real data are shown in black and the artificial star input grid is shown in red. The magnitude plot is centred on the white dwarf sequence.}
\label{asgrid}
\end{figure*}

First, 1000 values were chosen, evenly spanning the range of 18 to 28 magnitudes in F225W. 1000 values were also chosen, evenly spanning 16 to 26 in F336W. A radial coordinate grid was then defined in the drizzled image frame. This grid consists of concentric rings centred on the core of the cluster with a spacing of 50 pixels between each. Points on each annulus are evenly distributed with a 50 pixel spacing between adjacent artificial stars on the same annulus. This means there are approximately 25,000 points in the drizzled frame. This spacing ensures that the density of artificial stars has no effect on the measurements of the artificial stars. Although the artificial stars can not interfere with measurements of other artificial stars, the real stars can interfere with the measurements of artificial stars, which is what is measured by these tests. The locations of the artificial stars in magnitude and position space are shown in Figure \ref{asgrid}.

For each of the 1000 defined magnitudes in each of the two filters, one modified image was generated. Each of these images is identical to the original science image, but with the addition of 25,000 artificial objects, all of the same magnitude. So, in total, 2000 new WFC3 images were generated (1000 for each filter). This process was repeated to generate an additional 2000 new images for the ACS data.

Each of these 2000 images was run through the same reduction pipeline used to generate the catalogue of real objects. Whether or not the artificial object was found during the reduction was recorded. If the object was found, the output photometry was recorded as well. 

This output artificial star catalogue is the basis for the error and completeness model. We can index the input artificial F225W magnitudes with $i$, the input F336W magnitudes with $j$ and the input positions with $k$. The real catalogue only contains stars that are identified in both filters. Because the positions indexed by $k$ are consistent across both filters, we can consider any value $i$ and $j$ and access the output measurements of those two values at a consistent position $k$. While measuring the completeness of the artificial stars in this way we assess $i*j*k=2.5\times10^{10}$ unique stars in the data-space.

Considering only stars that are recovered, each unique combination $(i,j,k)$ also has a measured $\Delta$F225W and $\Delta$F336W (the difference between the magnitudes measured by the reduction pipeline and those input), as well as a $\Delta R$. From here on, the difference in input and output position is ignored ($\Delta R = 0$). Our analysis assumes that stars have no uncertainty in their position in the field. All of these data are then used to build a five dimensional array indexed by: F225W$_\mathrm{input}$, F336W$_\mathrm{input}$, $R$, $\Delta$F225W, and $\Delta$F336W.

\begin{figure}[htbp]
\centering
\epsfig{file=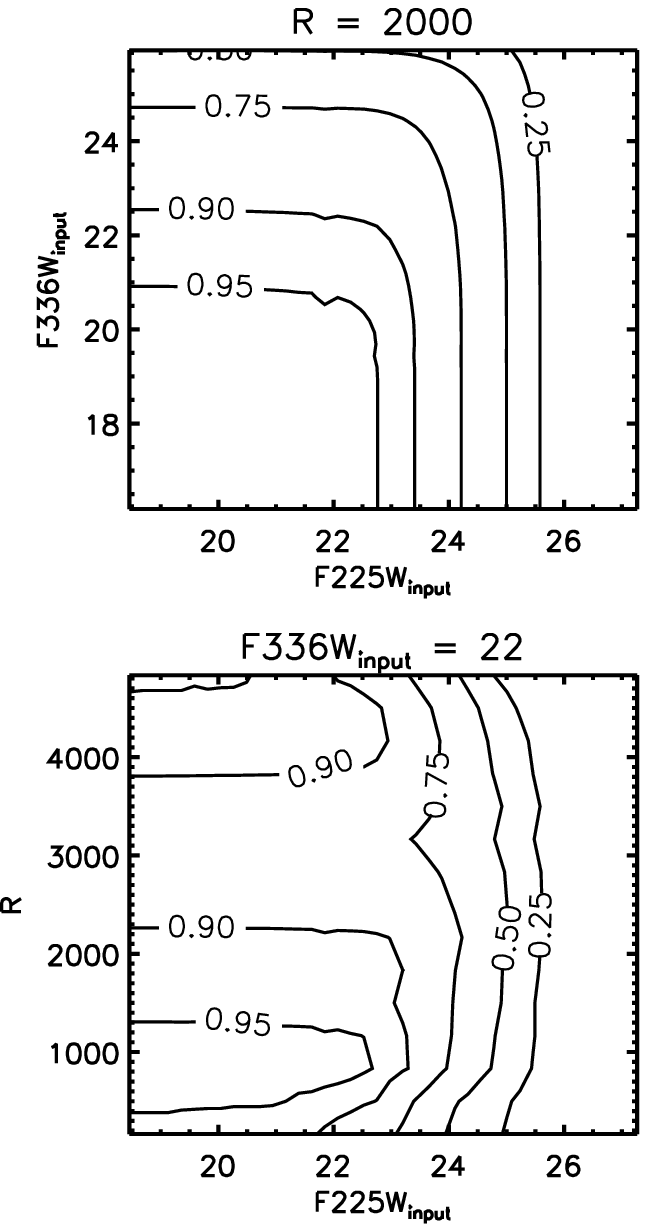,scale=1.3}
\caption{Completeness fraction in magnitude space for a fixed radial distance is shown on the top. Completeness fraction along the two-dimensional plane of $R$ and F225W$_\mathrm{input}$ with a fixed value of F336W$_\mathrm{input}$ is shown on the bottom. Radial distance is in units of WFC3 pixels, which correspond to 0.04 arc seconds each. These surfaces are the values from $C$ described in Equation \ref{compint}. $C$ is indexed by three parameters. In each plot above, one of these three is fixed, and $C$ is shown varying across the remaining two.}
\label{compslices}
\end{figure}

We call this array $E$, more explicitly: $E($F225W$_\mathrm{input},$F336W$_\mathrm{input},R,\Delta$F225W$,\Delta$F336W$)$. $E$ is generated by counting the artificial stars found after passing through our reduction procedure and dividing the count in each bin by the number of input artificial stars. The completeness of objects at a given pair of magnitudes and position in the field can be calculated as

\begin{equation} \label{compint}
C=\int_{-\infty}^{\infty} \int_{-\infty}^{\infty}E\ d(\Delta \mathrm{F225W})\ d(\Delta \mathrm{F336W}).
\end{equation}

How the completeness varies across the three-dimensional data-space is $C($F225W$_\mathrm{input},$F336W$_\mathrm{input},R)$. The completeness values on two orthogonal slices through this three dimensional space are shown in Figure \ref{compslices}. The odd structure in the bottom plot is real and not an effect of poor sampling. Our observational design leads to much larger total integration time in the core of the cluster than in the surrounding annuli, and this effect contributes positively to completeness as you move toward the core of the cluster. However, stellar density also increases dramatically toward the core, which decreases the completeness. These two effects together lead to the behaviour seen in the bottom figure.

\begin{figure}[htbp]
\centering
\epsfig{file=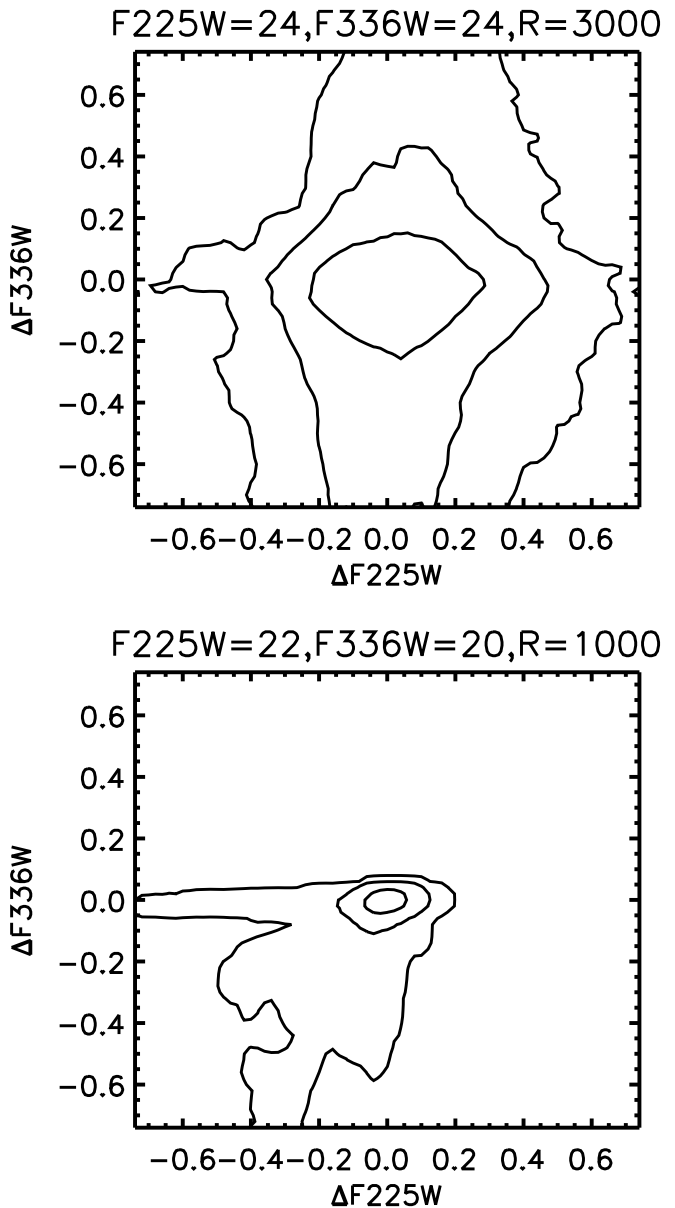,scale=1.3}
\caption{1, 2, and 3$\sigma$ contours (enclosing $68\%$, $95\%$, and $99.7\%$) of the photometric error distribution are shown at two points in the data-space. The top panel shows the error distribution for faint stars close to the outside of the inner WFC3 field. The lower panel shows the error distribution for brighter stars closer to the centre. R is again in units of pixels.}
\label{errslices}
\end{figure}

Another way to slice $E$ is to fix the first three parameters. The variations in $E$ over the remaining two parameters ($\Delta$F225W and $\Delta$F336W) are the error distribution of those input stars. Two such distributions are shown in Figure \ref{errslices}.

The error distribution describes the expected measured differences between the input and output artificial star photometry. From the figures one can see that the error distribution is highly non-Gaussian, and changes significantly across the space of the data. The five dimensional array $E$ is the key to transforming from predicted model magnitudes to the likelihood of actually measuring objects at various magnitudes in the catalogue. $E$ is an empirically determined model of our photometric scatter and measurement biases caused by the crowding of the field, the instrument and filters used, the observation pattern, and the reduction process. How this information is used in conjunction with our theoretical cooling models is described in Section \ref{modtrans}.

\subsection{MESA Models}
\label{mesa}

To calculate the cooling models we will compare to the data, we used Modules for Experiments in Stellar Astrophysics or MESA \citep{paxton2011modules}. We used these modules to perform simulations of stellar evolution starting with pre-main-sequence models with a metallicity of Z$=3.3\e{-3}$ appropriate for the cluster 47 Tuc. Wind parameters in the RGB and AGB phases of evolution are adjusted to produce white dwarfs with varying hydrogen layer thickness. The parameter adjusted is $\eta$. The mass loss rate on the RGB is calculated from \cite{reimers1975}:

\begin{equation} \label{rgbmassloss}
\dot{M}_\mathrm{RGB}=4\e{-13} \eta \frac{L}{L_{\odot}} \frac{R}{R_{\odot}} \frac{M_{\odot}}{M}.
\end{equation}

The mass loss rate on the AGB is calculated from \cite{bloecker1995}:

\begin{equation} \label{agbmassloss}
\dot{M}_\mathrm{AGB}=4.83\e{-9} \left( \frac{M}{M_{\odot}} \right)^{-2.1}  \left( \frac{L}{L_{\odot}} \right)^{2.7} \dot{M}_\mathrm{RGB}.
\end{equation}

Hydrogen layers with $q_\mathrm{H}$ between $10^{-6}$ and $10^{-3}$ are produced, where $q_\mathrm{H}$ is the fraction of the WD mass that is in the hydrogen layer. The thickness of the residual helium layer is $-1.45<\mathrm{log}_{10}(q_{He})<-1.30$ for all of the models that reach the white dwarf stage. Since we adjust the wind parameters and not the thickness of these layers directly, the masses of hydrogen and helium vary coincidentally as $\eta$ is changed. This process is stochastic, and very similar wind parameters can produce white dwarfs with hydrogen layers that differ by orders of magnitude. However, using this revision of MESA, it was not possible to produce helium layers that vary over such a range. The mass of the helium layer stays relatively constant over all of our simulations. As a result, we do not consider the helium layer mass as a separate parameter in our fitting. The fit distributions that we find can be considered as the result after marginalizing over this parameter.

During the evolution of these models, neutrino production is multiplied by a factor $A_{\nu}$ ranging from 0.1 to 3.18. We use SVN revision 5456 of MESA and start with the model \texttt{1M\_pre\_ms\_t\_wd} in the test suite. We changed the parameters \texttt{initial\_mass} and \texttt{initial\_z} of the star and adjusted the parameter \texttt{log\_L\_lower\_limit} to $-$6 so the simulation would run well into the WD cooling regime. We defined the starting time of our cooling models to be the last local maximum in luminosity. After the time we define as $t_0$, the models only decrease in brightness.

Initially it was thought that the mass of the white dwarf would be a dominant parameter. In fact, using the MESA framework, it is rather difficult to produce white dwarfs that differ significantly from the canonical globular cluster value of $0.53$M$_{\odot}$ by adjusting $\eta$. In order to change this mass by a large amount, the main sequence evolution time scale or the turn-off mass need to be adjusted to values that are entirely inconsistent with other observations in 47 Tuc and other clusters. Additionally, even when these models were produced, it was found that varying the mass does not produce a large effect on the predicted distribution of white dwarfs on our CMD. This is especially true when uncertainties in distance are considered, which are largely degenerate with uncertainties in white dwarf mass as seen on the CMD. For this reason, we have chosen to fix this parameter and instead look to the hydrogen layer thickness and the neutrino production as our fit parameters. Both of these have a much larger effect on the white dwarf cooling distribution than the mass does.

When producing these models, we found that most values of $\eta$ lead to white dwarfs with thicker atmospheres $(q_H>10^{-4})$, and that the mass of the resulting hydrogen layer is independent of $\eta$ for most values of that parameter. \cite{althaus2015white} find similarly that the mass of H with which the white dwarf enters its cooling track is not altered by the occurrence of mass loss during these stages. The values of $\eta$ which are able to produce very thin atmospheres constitute a small portion of those considered. So, we produce a large group of models with varying $\eta$ and $A_{\nu}$ values. The white dwarfs that result from these models are then reparametrized over a grid of $q_{H}$ and $A_{\nu}$. The model grid in this space that we end up fitting to our data does not uniformly represent $\eta$. The intention is to provide good coverage over $q_{H}$ and $A_{\nu}$, since these are the parameters we are interested in studying. The cooling models in this parameter space are the starting point for the approach described in the next section.
 
\subsection{Moving the Model to Data-Space}
\label{modtrans}

The result of these calculations is a grid of white dwarf cooling sequences over varying $q_\mathrm{H}$ and $A_{\nu}$. Each combination of hydrogen layer thickness and neutrino factor produces one cooling sequence that describes how the physical properties of the white dwarf change with time. The ones we are interested in are effective surface temperature $T$, surface gravity $g$, and radius $r$. These three parameters can be thought of as functions of time $t$ but also $q_\mathrm{H}$ and $A_{\nu}$. The state of the model resulting directly from calculations in MESA is

\begin{subequations} \label{modeqs1}
\begin{align}
T(t,q_\mathrm{H},A_{\nu}), \\
g(t,q_\mathrm{H},A_{\nu}), \\
r(t,q_\mathrm{H},A_{\nu}).
\end{align}
\end{subequations}

These parameters exist in theory-space and need to be transformed to quantities in data-space. The first step is to determine how the model objects look through the HST filters. To do this we use spectral models for DA white dwarfs described in \citet{tremblay2011specda} and the publicly available HST filter throughputs available from \citet{stsci2009web}. We also model the reddening using the interstellar reddening model from \citet{fitzpatrick1999correcting}.

The spectral model grid provides the spectral flux density (energy per unit time per unit wavelength), $F$, at the surface of the object as a function of $T$ and $g$. Explicitly we can write this as $F(\lambda,T,g)$. 

The Fitzpatrick reddening curve is parametrized by $E(B-V)$, and $R_V$. These are defined as follows. $A_V$ is the extinction in $V$ and $A_B$ is the extinction in $B$ (the fraction of power lost in filter $V$ and filter $B$ respectively). $E(B-V)=A_B-A_V$; and $R_V=\frac{A_V}{A_B-A_V}$. The symbols $B$ and $V$ refer to magnitudes in the $B$ and $V$ filters (centred at 445nm and 551nm). We fix $R_V=3.1$ which is found to apply in most galactic environments \citep{fitzpatrick1999correcting}. This value also gives good agreement between our white dwarf model sequence and our data, suggesting that it is appropriate for 47 Tuc as well. We do not vary $R_V$, and it remains fixed at the standard value of 3.1 for the remainder of our analysis. The reddening curve is called $k(\lambda,e)$, and is the fraction of the power transmitted as a function of wavelength between the object and the observer.

The filter throughput curves describe the fractional power transmittance as a function of wavelength for a given filter. This will be labeled as $S_i(\lambda)$ where $i$ is an arbitrary index to indicate which filter is being used.

All of these objects are then combined as shown in Equation \ref{magcalc} to generate a predicted magnitude:

\begin{equation} \label{magcalc}
m_i'(T,g,r,e,d)=-2.5\mathrm{log}_{10}\left( \frac{\int_{0}^{\infty} \lambda F S_i k d\lambda}{\int_{0}^{\infty} \lambda F_0 S_i d\lambda} \right )+5\mathrm{log}_{10}\left(\frac{d}{r} \right ).
\end{equation}

Here $F_0$ is the zero-point flux density appropriate for each filter (the flux density that corresponds to a magnitude of zero). The new parameter $d$ is introduced as the distance to the cluster. If we substitute Equations \ref{modeqs1} into \ref{magcalc}, we get

\begin{equation} \label{modeqs2}
m_i'(t,q_\mathrm{H},A_{\nu},e,d). 
\end{equation}

At this point $m'$ is used to describe the magnitude rather than $m$, as the photometric biases described in Section \ref{arts} have not been taken into account yet. The two filters used in the WFC3 observations are F225W and F336W. Magnitudes in these filters will be referred to as $m_2$ and $m_3$, respectively.

The thing we wish to calculate is the probability density, $f$, of finding a white dwarf at a given set of magnitudes ($M_2$ and $M_3$) and a given position in the field ($R$). This is $f=\frac{dN}{dm_2 dm_3 dR}$. In order to manipulate Equation \ref{modeqs2} into the form we need to compare to the data, we first solve for $t$ as a function of everything else, and then take two derivatives:

\begin{equation} \label{lfunc1}
\frac{dt}{dm_2'dm_3'}=\frac{dt}{d(m_2')^2}\frac{dm_2'}{dm_3'}.
\end{equation} 

\vspace{12pt}

Next we multiply by the formation rate of objects in the field, $\dot N=\frac{dN}{dt}$:

\vspace{12pt}

\begin{equation} \label{lfunc2}
\frac{dN}{dm_2'dm_3'}=\frac{dt}{d(m_2')^2}\frac{dm_2'}{dm_3'}\frac{dN}{dt}.
\end{equation}

\vspace{12pt}

Now we have to bring back the error distribution, $E$, from Section \ref{arts}. Before we refered to this as $E($F225W$_\mathrm{input},$F336W$_\mathrm{input},R,\Delta$F225W$,\Delta$F336W$)$, but now we can see that F225W$_\mathrm{input}=m_2'$, F336W$_\mathrm{input}=m_3'$, $\Delta$F225W$=m_2-m_2'$, and $\Delta$F336W$=m_3-m_3'$. So what we have is:

\begin{equation}
E(m_2-m_2',m_3-m_3',m_2',m_3',R).
\end{equation}

We can perform an integral to transform Equation \ref{lfunc2} into the form that we want, which is $f$:

\begin{widetext}
\begin{equation} \label{convint}
f=\frac{dN}{dm_2dm_3dR}=\rho(R)\int_{-\infty}^{\infty} \int_{-\infty}^{\infty}  \frac{dN}{dm_2'dm_3'}E(m_2-m_2',m_3-m_3',m_2',m_3',R)dm_2'dm_3'.
\end{equation}
\end{widetext}

This step is a convolution integral of the error distribution with the theoretical two dimensional luminosity function. The error distribution changes with position in data-space. The assumed radial density distribution, $\rho(R)$, must also be included at this point. Neglecting this term is equivalent to an assumption of a density distribution that is constant with radius. We use a King-Michie model \citep{king-1966-aj,michie-1963-mnras} to calculate the projected density distribution. The assumed King radius and tidal radius values are taken from \cite{goldsbury2013quantifying} and \cite{harris-1996-aj}, respectively ($r_0=32''$ and $r_t=3800''$). Note that $\rho(R)$ in Equation \ref{convint} is not the three dimensional density distribution, it is the projected and azimuthally integrated distribution with units of objects per pixel.

At this point we have a distribution $f$ that gives the probability density of finding an object at any point in the data-space $(F225W,F336W,R)$ for any combination of our model parameters $(q_\mathrm{H},A_{\nu},d,e,\dot N)$. We will use a test statistic referred to as the ``unbinned likelihood" to determine how well the data are predicted by this model. 

\subsection{The Unbinned Likelihood}

A quick derivation of the unbinned likelihood starts by considering a bin in the data-space with dimensions $\Delta R$ by $\Delta m_2$ by $\Delta m_3$. This bin is centred at at the location $(R_j,m_{2k},m_{3l})$. From the model discussed above, the predicted number of objects in this bin is given as $f(R_j,m_{2k},m_{3l})\Delta R \Delta m_2 \Delta m_3$. The probability of finding $n$ objects in this bin is given by the Poisson distribution:

\begin{widetext}
\begin{equation}	\label{poisprob}
P(n;f(R_j,m_{2k},m_{3l}))=\frac{[f(R_j,m_{2k},m_{3l})\Delta R \Delta m_2 \Delta m_3]^ne^{-f(R_j,m_{2k},m_{3l})\Delta R \Delta m_2 \Delta m_3}}{n!} .
\end{equation}

The limit is then taken as the bin size approaches zero. This leaves a number of bins equal to the number of data points with one object in them, and an infinite number of bins with zero objects in them. Then there are two possible forms for the probability to take:

\begin{subequations} \label{twocases}
\begin{align}
P(1;f(R_j,m_{2k},m_{3l}))&=f(R_j,m_{2k},m_{3l})\Delta R \Delta m_2 \Delta m_3 e^{-f(R_j,m_{2k},m_{3l})\Delta R \Delta m_2 \Delta m_3}, \\
P(0;f(R_j,m_{2k},m_{3l}))&=e^{-f(R_j,m_{2k},m_{3l})\Delta R \Delta m_2 \Delta m_3}.
\end{align}
\end{subequations}

The log-likelihood is the sum of the logarithm of the probabilities of all of the bins:

\begin{equation} \label{loglsum}
\mathrm{log} L=\sum_{i}^{ } \mathrm{log}(f(R_i,m_{2i},m_{3i}))+\sum_{i}^{ } \mathrm{log}(\Delta R \Delta m_2 \Delta m_3)-\sum_{jkl}^{ }f(R_j,m_{2k},m_{3l})\Delta R \Delta m_2 \Delta m_3.
\end{equation}
\end{widetext}

The first two sums are only performed over $i$, which is the index of the data points, since these terms only exist in the case of Equation \ref{twocases}a. The third sum is carried out for all bins, even those with no data in them, since this term is common to Equations \ref{twocases}a and \ref{twocases}b. The second sum does not depend in any way on the data or our chosen model. It will be an additive constant to the log-likelihood and so can be ignored. Given the limit as the bins become infinitesimal, we can also see that the third sum becomes an integral of the probability density over the entire data-space. With these two modifications we have

\begin{equation} \label{likeint}
\mathrm{log} L=\sum_{i}^{ } \mathrm{log}(f(R_i,m_{2i},m_{3i}))-\iiint \limits_{\mathrm{data-space}} f dR dm_2 dm_3 ,
\end{equation}

\begin{equation} \label{likenpred}
\mathrm{log} L=\sum_{i}^{ } \mathrm{log}(f(R_i,m_{2i},m_{3i}))-N_\mathrm{pred}.
\end{equation}

The integral in Equation \ref{likeint} is the total number of objects that our model predicts in the region of data-space in which we are doing the comparison. To simplify the notation, we have written this as $N_\mathrm{pred}$ in Equation \ref{likenpred}. The bound region in data-space does not have to be rectangular or any other particular shape; as long as the integral in Equation \ref{likeint} is performed over exactly the same region from which we select the data, this works.

The description up  until this point has been specific to the central WFC3 data, but a virtually identical approach is used for the ACS data. All nine ACS fields were reduced separately. Since we assume that the cluster distribution and also the incompleteness are azimuthally symmetric, we only perform artificial star tests on one of these nine images. These tests are still parametrized by radial position $R$ and two magnitudes (in this case $F435W$ and $F555W$). We can use these tests to build a second probability density function $f$ specific to the ACS data.

Although it is left out of the likelihood equations above, the probability density functions $f$, and therefore the likelihood as well, both depend on the choice of model parameters $(q_\mathrm{H},A_{\nu},d,e,\dot N)$. The first four parameters are consistent across both the ACS data and the WFC3 data. This is because all of the stars should have the same hydrogen layer thickness, have the same neutrino physics, lie at the same distance, and experience the same reddening. The fifth parameter $\dot N$ is the rate at which white dwarfs are being produced in each field, so will not be the same for both data sets. 

\begin{figure*}[!htbp]
\centering
\epsfig{file=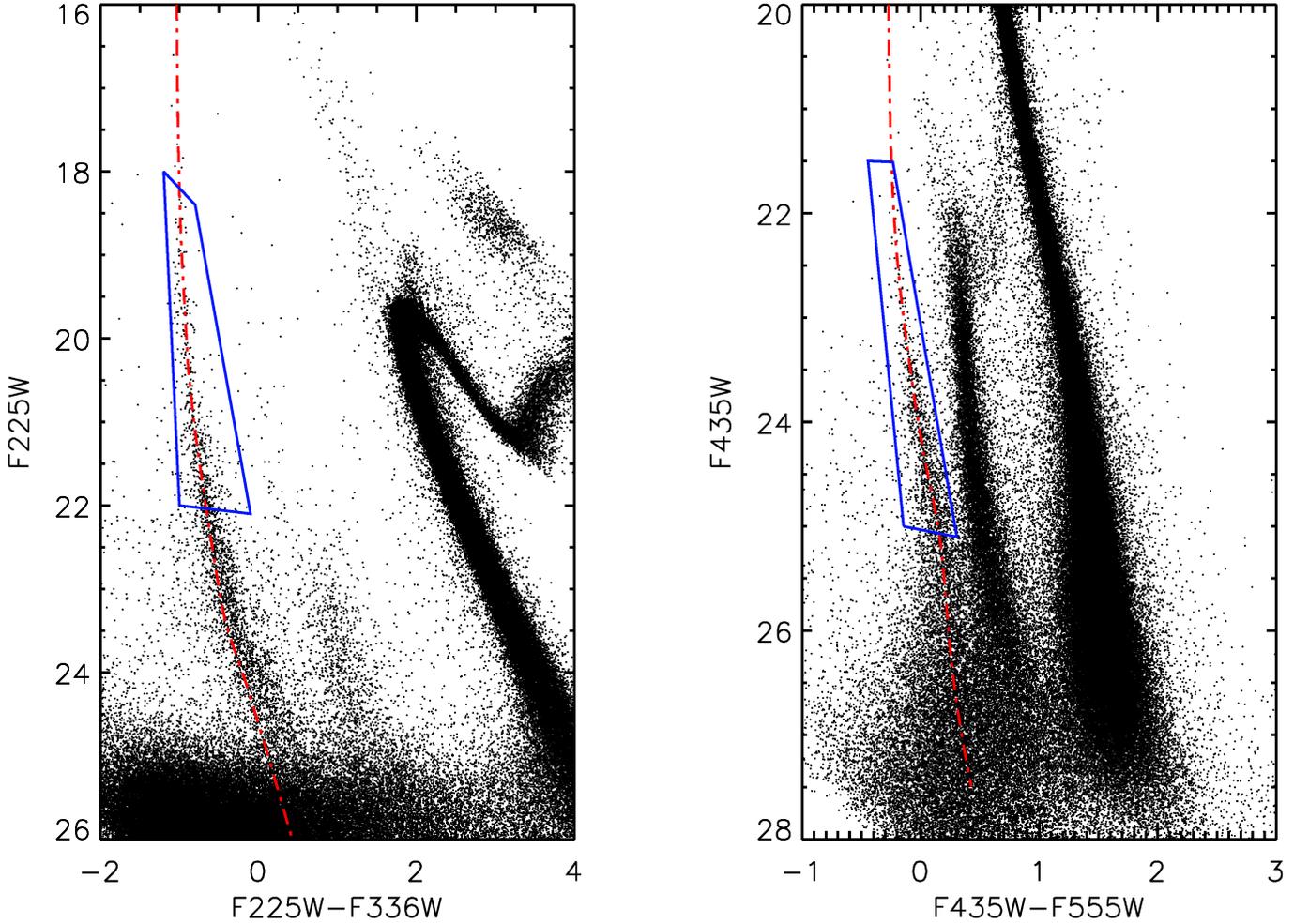,width=\linewidth}
\caption{The colour magnitude diagrams of the two fields. Overplotted are the region in which we fit our model in blue, and the white dwarf cooling sequence in red for the mean prior values given in Table \ref{priors}, before convolution with the error model.}
\label{fitreg}
\end{figure*}

To understand the $\dot N$ parameter more intuitively, imagine first selecting the region shown in Figure \ref{fitreg}. If we could watch this CMD change over hundreds of millions of years, we would see objects moving down the white dwarf sequence as they cool and leaving the region through the bottom. We would also see new white dwarfs arriving through the top of the region as they contract and cool coming from the horizontal branch. $\dot N$ would be the flux into the fit region as observed in this way. This flux changes on long timescales, as the cluster turn-off moves down the mass function to lower mass stars, but it will change slowly enough over the age range of white dwarfs in the fit region that we can assume that it is constant. 

In total we now have six parameters; $(q_\mathrm{H},A_{\nu},d,e,\dot N_\mathrm{WFC3}, \dot N_\mathrm{ACS})$. We can combine the two likelihood functions to generate a likelihood over parameter space for all of the data together.

\subsection{Degenerate Parameters and Prior Knowledge}

A number of the parameter constraints are partially degenerate. Distance and reddening are partially degenerate, due to the fact that a larger reddening requires a larger extinction in each filter, producing a similar effect to putting the cluster farther away. The formation rate is degenerate with distance, which is because the cooling rate roughly follows a power law with time. For a perfect power law it is impossible to distinguish between a vertical and horizontal shift in log-space. Because the model cooling rate is similar, changing the timescale looks roughly the same as making all of the objects brighter or fainter. 

In addition to these degeneracies, the data set is not particularly useful for obtaining constraints on these parameters. Without using any prior distributions on other parameters, the fit distributions for distance or reddening are 2-3 times wider than current constraints. This isn't worrying however, since the parameters we are most interested in studying are $q_\mathrm{H}$ and $A_{\nu}$, and the data are able to provide useful constraints on these parameters. So for these reasons, we choose to use priors for all of our model parameters except for $q_\mathrm{H}$ and $A_{\nu}$. This leads to a constraint on the neutrino production rate and the mass of the hydrogen layer, dependent on what we already know about the other cluster parameters.

\begin{table}[h]
\caption{Values for $\dot N$ are determined as described in the paragraph following this table. They are provided in units of objects per Myr.}

\centering
\begin{tabular}{|l || c | c | c |}

\hline
Parameter & Mean & Std. Dev. & Reference\\
\hline

$(m-M)_0$ & 13.30 & 0.13 & \cite{woodley2012spectral}\\
$E(B-V)$ & 0.04 & 0.02 & \cite{harris-1996-aj}\\
$\dot N_\mathrm{WFC3}\ ($Myr$^{-1})$ & 8.2 & 0.3 & \\
$\dot N_\mathrm{ACS}\ ($Myr$^{-1})$ & 2.61 & 0.07 & \\

\hline
\end{tabular}
\label{priors}
\end{table}

We use Gaussians for all of the prior distributions. The parameters of these are shown in Table \ref{priors}. The values for the formation rate of white dwarfs in each field $(\dot N)$ are estimated in the same way as in \cite{goldsbury2012empirical}. We use data on the giant branch in each field, along with completeness corrections from the artificial star tests, and main sequence evolution models from \cite{dotter2008dartmouth} to model the rate at which stars are leaving the main sequence and evolving up the giant branch. We make the assumption that this is the same rate at which stars are arriving onto the white dwarf sequence. 

\subsection{Restricting the Fitting Region}

The region in magnitude space over which we fit our model is restricted in both fields to the region immediately around the white dwarf sequence. These regions are shown in blue in Figure \ref{fitreg}. There are two main reasons for the requirement that the fitting be restricted to these regions. In both the WFC3 and ACS data, the lower white dwarf sequence begins to overlap with the main sequence of the Small Magellanic Cloud, which lies in the background of 47 Tuc. 

We do not include the SMC in our model and so fitting over this region would be unreliable. In the case of the WFC3 data, the white dwarf models also deviate significantly from the measured sequence at fainter magnitudes. The cause of this discrepancy is unclear. After a thorough analysis, we find no apparent systematic offset introduced by the photometric reduction process. The photometry of the sequence is slightly biased by diffuse background light and crowding of the inner field, but this could account for only $10\%$ of the discrepancy observed at most. Additionally, the error model described in Section \ref{arts} fully accounts for the discrepancy attributable to diffuse background light. The model offset even persists through entirely different reductions performed with and without drizzling. Avoiding the lower WFC3 sequence does have a considerable impact on the total sample size, but if the fitting region were not restricted as shown in Figure \ref{fitreg}, all of the white dwarf models would be strongly ruled out. Using the region shown, there are regions of parameter space that are consistent with the data. Although the plot shows this region in colour-magnitude space, fitting is done in magnitude-magnitude space.

\begin{figure}[!htbp]
\centering
\epsfig{file=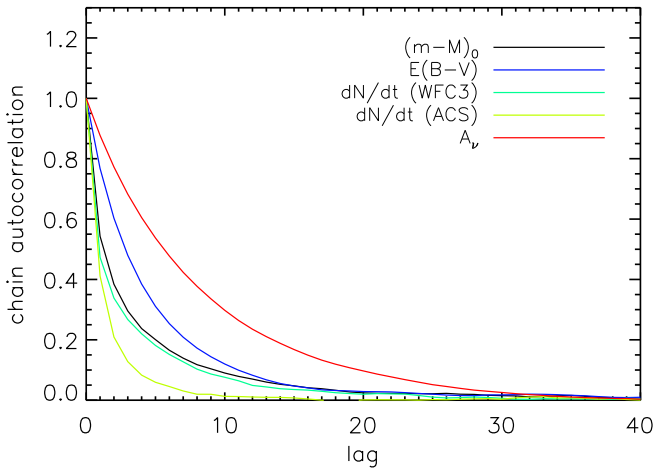,width=\linewidth}
\caption{The auto-correlation of each chain parameter as a function of the lag between points in the chain. No memory of the past position in parameter space is retained beyond 40 steps.}
\label{aclag}
\end{figure}

\begin{figure}[!htbp]
\centering
\epsfig{file=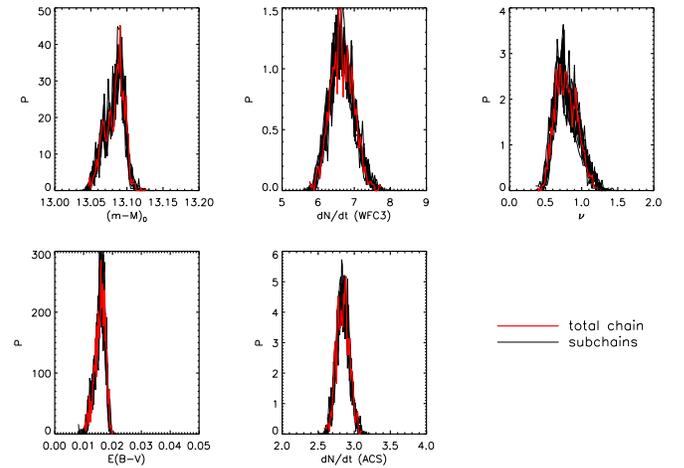,width=\linewidth}
\caption{A comparison of the likelihood distribution for each of our model parameters shown for the entire chain in red. The thinner black lines are independent segments 1000 steps long. Each of these separate chains is much longer than the length scale on which correlations persist in the chains. All of them average to the same broad distribution that we find with the entire chain, but with more statistical noise.}
\label{subchains}
\end{figure}

\subsection{Markov Chain Monte Carlo Sampling}

To explore the model parameter space, we use a Markov chain Monte Carlo (MCMC) method. Specifically, we use the Metropolis-Hastings algorithm \citep{metropolis1953equation,hastings1970monte} with Gibbs sampling \citep{geman1984stochastic}. The total chain used to generate the parameter distributions in this analysis is 100,000 points long. The process of thinning, in which only every $N$th point in a Markov chain is kept, is often used to reduce the correlation between points within the chain. We do not follow this approach, since these small-scale correlations are averaged out by the comparably long length of the total chain. This is explained in detail in \cite{link2012thinning}. We can look at the auto-correlation of each of our model parameters to see that beyond 20 steps the auto-correlation drops to effectively zero (Figure \ref{aclag}). It is possible to break the total chain into many smaller chains, each of which gives the same result within statistical uncertainties (Figure \ref{subchains}). These simple tests indicate that the total chain will be not be biased by the small length scale correlations within it.

\section{Results}

\subsection{Hydrogen Layer Thickness ($q_\mathrm{H}$) and Neutrino scaling ($A_{\nu}$)}

\begin{figure}[!htbp]
\centering
\epsfig{file=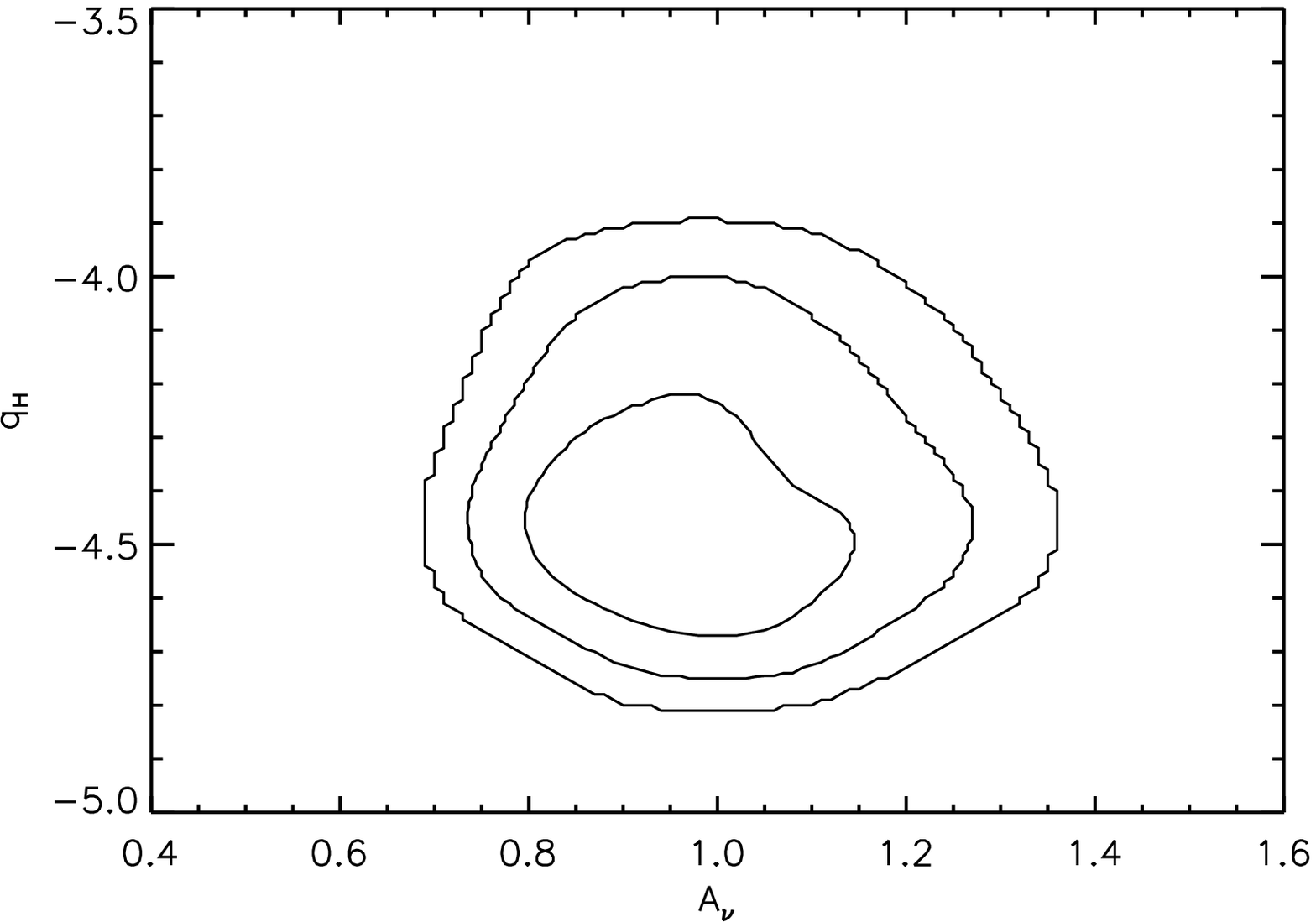,width=\linewidth}
\caption{Contours of the join fit distribution of $A_{\nu}$ and $q_\mathrm{H}$. 1, 2, and 3$\sigma$ contours (enclosing $68\%$, $95\%$, and $99.7\%$) are shown.}
\label{qhnufit}
\end{figure}

\begin{figure}[!htbp]
\centering
\epsfig{file=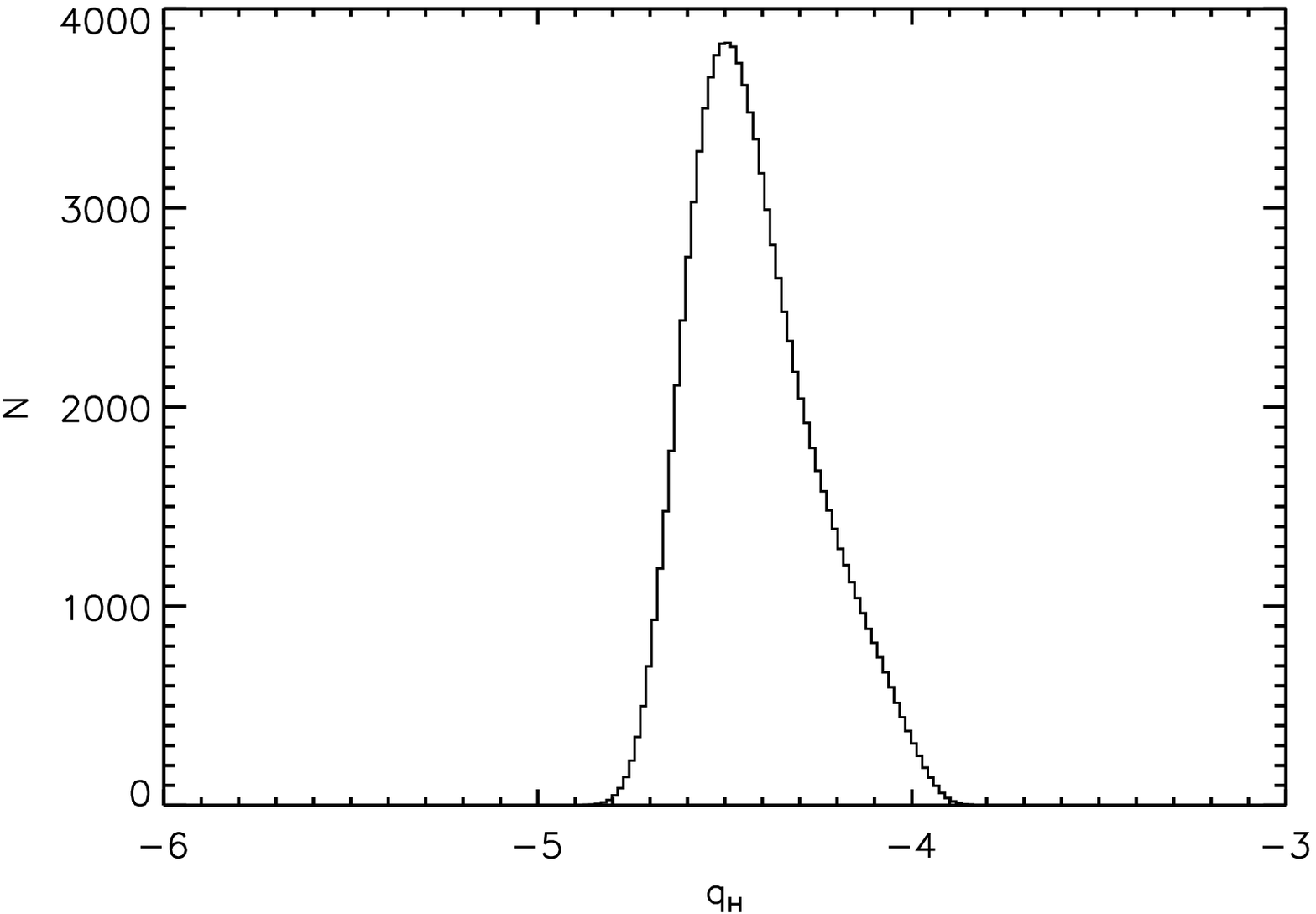,width=\linewidth}
\caption{A histogram of the posterior distribution of $q_\mathrm{H}$ values from the markov chain. This is the likelihood distribution of $q_\mathrm{H}$ after marginalizing over all other parameters in the model with prior distributions on each, except for $A_{\nu}$. The range of $2.2\e{-5}$ to $8.9\e{-5}$ contains $95\%$ of the distribution.}
\label{qhfit}
\end{figure}

\begin{figure}[!htbp]
\centering
\epsfig{file=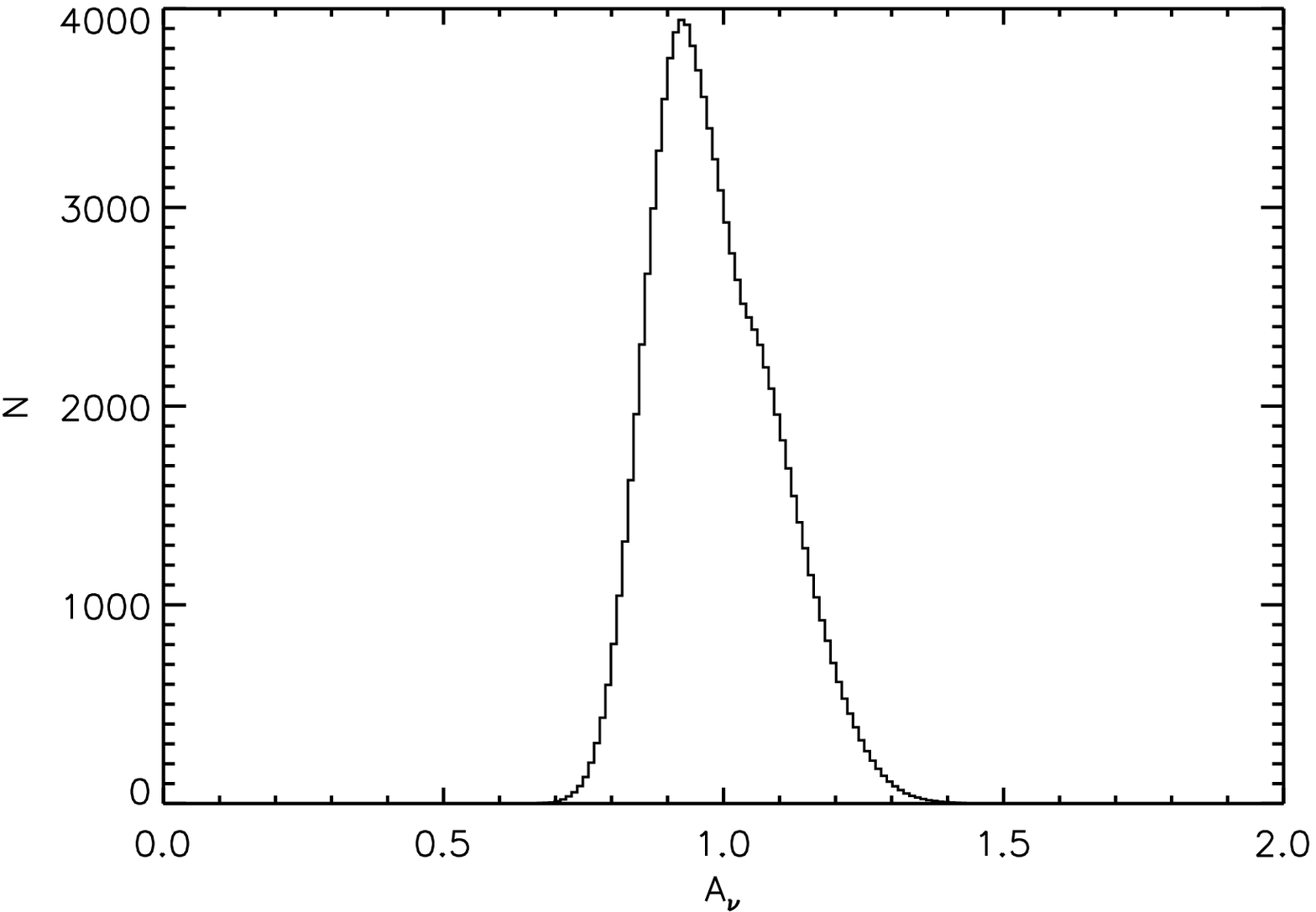,width=\linewidth}
\caption{A histogram of the posterior distribution of $A_{\nu}$ factor values from the markov chain. This is the likelihood distribution of $A_{\nu}$ after marginalizing over all other parameters in the model with prior distributions on each, except $q_\mathrm{H}$. The range of $0.83$ to $1.22$ contains $95\%$ of the $A_{\nu}$ fit distribution.}
\label{nufit}
\end{figure}

The only two parameters fit without priors are $A_{\nu}$ and $q_\mathrm{H}$. The join fit distribution for these parameters is shown in Figure \ref{qhnufit}. The distribution for $q_\mathrm{H}$ alone is shown in Figure \ref{qhfit}. This is the histogram of the $q_\mathrm{H}$ values in the total chain. It is also the likelihood distribution of this parameter after marginalizing over all of the other parameters. The range of $2.2\e{-5}$ to $8.9\e{-5}$ contains $95\%$ of the distribution. The distribution for $A_{\nu}$ alone is shown in Figure \ref{nufit}. This is the histogram of $A_{\nu}$ values in the total chain. It is also the likelihood distribution of this parameter after marginalizing over all other parameters. The range of 0.83 to 1.22 contains $95\%$ of the distribution. The default amount of neutrino production in the MESA models corresponds to $A_{\nu}=1$. Our findings are consistent with this. 

\begin{figure}[!htbp]
\centering
\epsfig{file=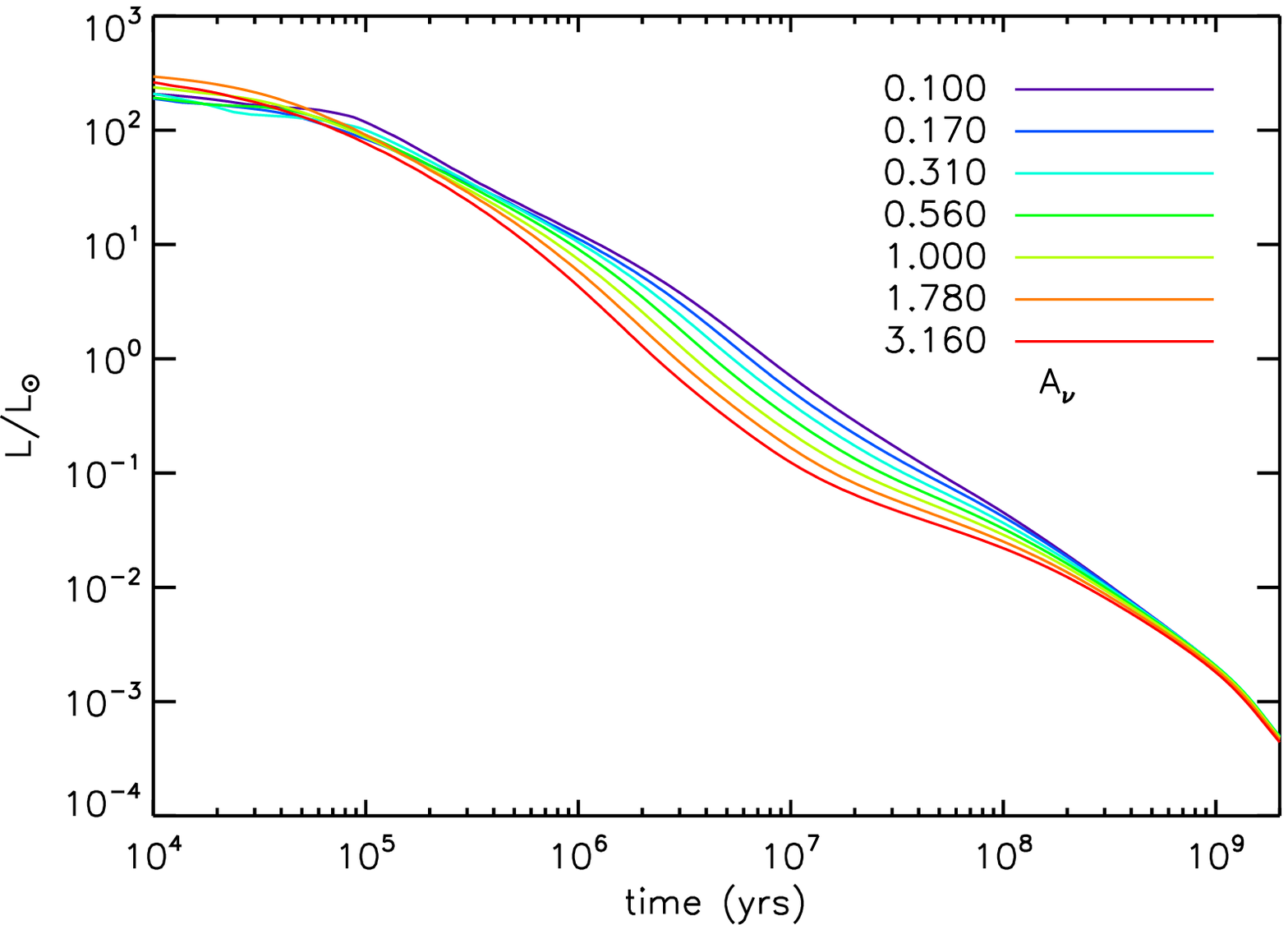,width=\linewidth}
\caption{The relation between luminosity and time shown for varying $A_{\nu}$ values while $q_\mathrm{H}$ is held fixed.}
\label{mesanvary}
\end{figure}

\begin{figure}[!htbp]
\centering
\epsfig{file=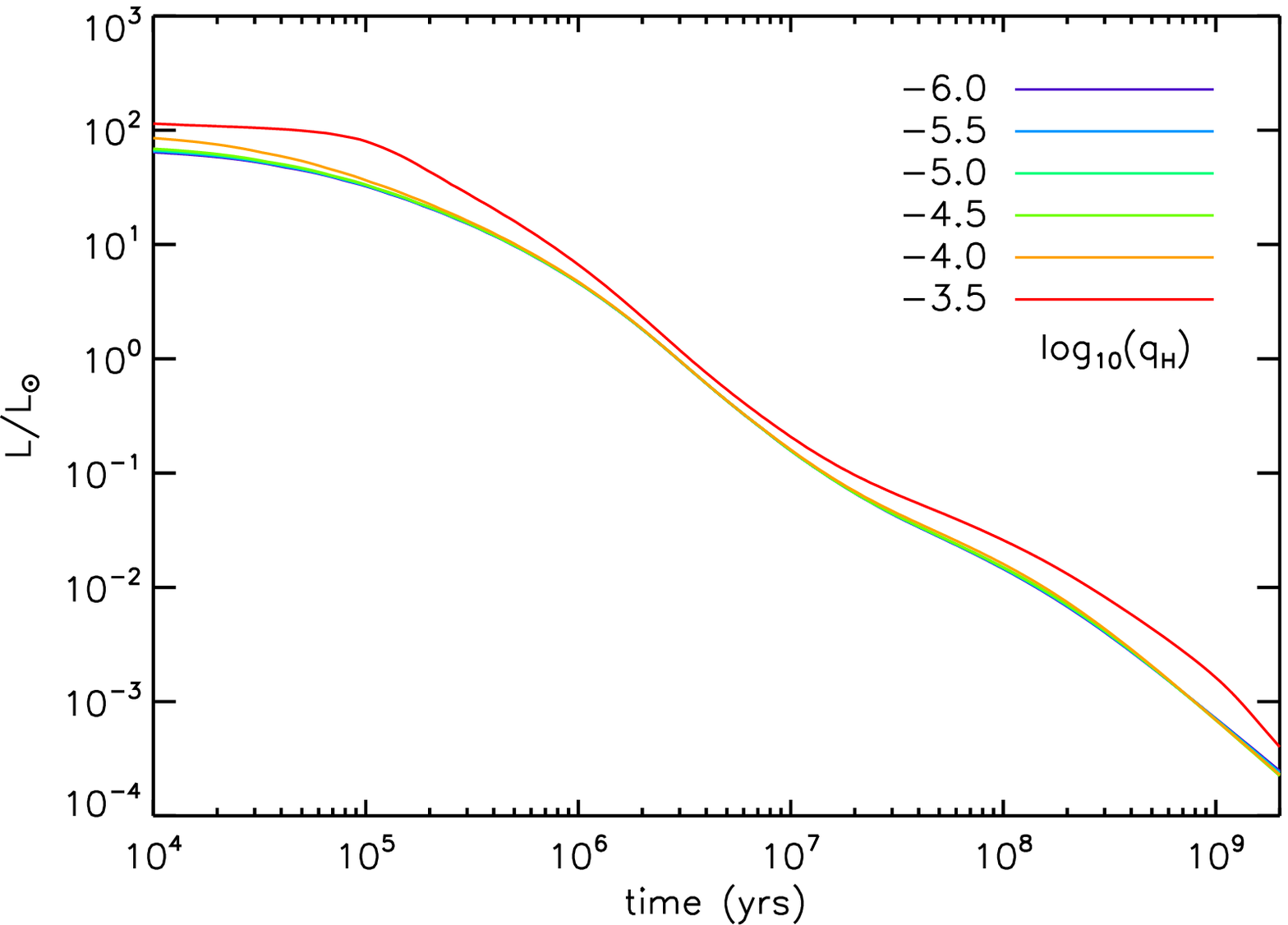,width=\linewidth}
\caption{The relation between luminosity and time shown for varying $q_\mathrm{H}$ values while $A_{\nu}$ is held fixed.}
\label{mesaqhvary}
\end{figure}

Figures \ref{mesanvary} and \ref{mesaqhvary} show how the cooling models change with varying $A_{\nu}$ and $q_\mathrm{H}$ respectively. From these figures it is easy to see that changing $A_{\nu}$ has a much larger effect than changing $q_\mathrm{H}$ over the range of values we are considering. This results in a tighter constraint on $A_{\nu}$.

\subsection{Neutrino Species}

The MESA software uses the results of \cite{itoh1996neutrino} to calculate the neutrino emissivity. The neutrino production in white dwarfs in this temperature range is dominated by the plasmon neutrino process. The term ``plasmon" refers to a photon moving through a plasma. Unlike photons in a vacuum, plasmons travel more slowly than $c$ due to their interaction with the plasma. Some of the energy of the photon is transferred to the free electrons in the plasma. This allows the interacting plasmon to decay into a neutrino-antineutrino pair while still conserving momentum: $\gamma \rightarrow \nu \ + \ \bar{\nu}$ \citep{kantor2007neutrino}. The factor we refer to as $A_{\nu}$ is a multiplication of $Q_\mathrm{plasma}$ from equation 4.1 of \cite{itoh1996neutrino}. The plasmon neutrino rate is calculated in the MESA code as described in \cite{itoh1996neutrino}, and this rate depends on the central density of the white dwarf, core temperature of the white dwarf, the weak mixing angle (more often parametrized as $\mathrm{sin}^2\theta_\mathrm{W}$), and the effective number of non-electron neutrino species ($n$). We multiply the rate that comes out of this calculation by the factor $A_{\nu}$. The rate $Q_\mathrm{plasma}$ can be affected by any of the four parameters listed above, and so the factor $A_{\nu}$ could be thought of as a change to any combination of these four. However, interpreting $A_{\nu}$ as a change to either the central density or temperature would be inconsistent, since these parameters were not consistently altered elsewhere in the code. It therefore only makes sense to interpret a constraint on this parameter as either a constraint on $\mathrm{sin}^2\theta_\mathrm{W}$ and $n$. 

The relation between $A_{\nu}$, $\mathrm{sin}^2\theta_\mathrm{W}$, and $n$ is 

\begin{equation} \label{neut2deq}
A_{\nu} \propto (C_V^2+nC_V'^2),
\end{equation}

where $C_V$ and $C_V'$ are defined as

\begin{equation} \label{cv}
C_V=\frac{1}{2}+2\mathrm{sin}^2\theta_\mathrm{W}
\end{equation}

and

\begin{equation} \label{cv'}
C_V'=1-C_V.
\end{equation}

\begin{figure}[!htbp]
\centering
\epsfig{file=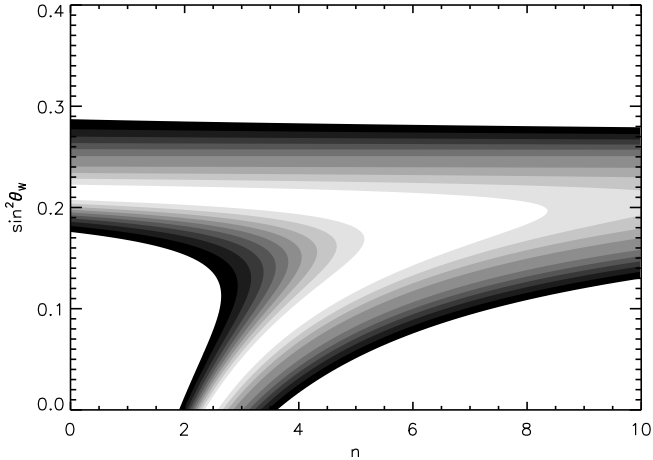,width=\linewidth}
\caption{The likelihood of $\mathrm{sin}^2\theta_\mathrm{W}$ and $n$. For the accepted value of $\mathrm{sin}^2\theta_\mathrm{W}$ (0.23155 from \cite{PDG}), $C_V'$ is very small. This means that the dependence of $A_{\nu}$ on $n$ is very weak. As a result, our constraint on the neutrino production ($A_{\nu}$) does not reasonably constrain $n$.}
\label{weinconts}
\end{figure}

Using these relations, the likelihood in Figure \ref{nufit} can be mapped onto the two-dimensional space of $\mathrm{sin}^2\theta_\mathrm{W}$ and $n$. This is shown in Figure \ref{weinconts}. From the Particle Data Group \citep{PDG} $\mathrm{sin}^2\theta_\mathrm{W}=0.23155$. Looking at this value in Figure \ref{weinconts}, one can see that changing $n$ will not produced a large change in $A_{\nu}$. This means that our constraint on $A_{\nu}$ does not allow us to put a reasonable constraint on $n$.

\subsection{Fit Quality}

While these distributions in parameter space inform us which values lead to better or worse fits, they cannot indicate which fits are actually consistent with the dataset. This is possible in the same way that one could determine a best-fitting line for data that are clearly not linear. A maximum in likelihood will be found regardless of whether or not the fit is good. It is possible that even though we have found constrained likelihood distributions for $q_\mathrm{H}$ and $A_{\nu}$, the best fitting model would still be a poor fit to the data.

\begin{figure}[!htbp]
\centering
\epsfig{file=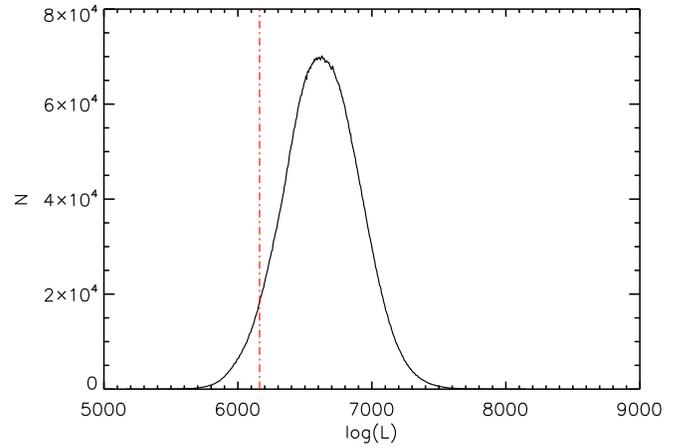,width=\linewidth}
\caption{The distribution of likelihood values expected when sample data are drawn from the best fitting cooling model is shown as the black distribution. The red line indicates the log-likelihood value found when comparing the real data to the model. We would expect to draw data from our best fitting model that differs at least as much as the real data do from that model 4.9$\%$ of the time.}
\label{fitqual}
\end{figure}

To evaluate whether this is the case we use our best-fitting model to create simulated data through Monte Carlo sampling. The model being referred to here is Equation \ref{convint}, so these simulated samples take full account of the empirically determined incompleteness as well as photometric bias and scatter. We generate millions of simulated data sets from the best-fitting model and in each case calculate the unbinned likelihood of the simulated data against the model. These values make up the distribution of likelihood we would expect if data such as ours were drawn from our best-fitting model. We can then ask whether or not the likelihood value we find from comparing the real data to the model is an outlier in this distribution. If the real likelihood value falls well outside the distribution of likelihood values from the simulated data, then we would conclude that it is unlikely the real data could be drawn from our best-fitting model. The distribution of likelihood values from simulated samples, as well as the likelihood value of the real data compared to the model, are shown in Figure \ref{fitqual}.

Integrating from the red dashed line to the left tail of the distribution in Figure \ref{fitqual} will enclose 4.9$\%$ of the total area under the black line. This indicates that  4.9$\%$ of the time we would expect to draw data from our model that differs at least as much as the real data do. 

\subsection{Cooling Models From Other Groups}
\label{othergroups}

We compared the four cooling models used in \cite{goldsbury2012empirical} to this data set as well. These models come from \cite{fontaine2001potential}, \cite{hansen2015constraining}, \cite{lawlor2006mass}, and \cite{renedo2010new}. The model from \cite{hansen2015constraining} is more recent than the version fit in \cite{goldsbury2012empirical}, while the other models are the same. These models do not have any intrinsic parameters to be varied, so the fits are only performed over the parameter space (with priors) defined by the four parameters in Table \ref{priors}. The quality of fit assessment for each model is done in the same way as for our MESA models (summarized in Figure \ref{fitqual}. Simulated data is repeatedly drawn from each best-fitting model and compared back to that model. The distribution of resulting likelihood values is then integrated from the value obtained when comparing the real data to the model. The results are shown in Table \ref{4mod_fitqual}. With this data set we cannot strongly rule out any of these models. 

\begin{table}[h]
\caption{The ``fit probability" is defined as the probability that data drawn from the given model would produce a likelihood value less than the real data do when compared back to that model.}

\centering
\begin{tabular}{|l || c |}

\hline
cooling model & fit probability\\
\hline

\cite{fontaine2001potential} & 0.032 \\
\cite{hansen2015constraining} & 0.025 \\
\cite{lawlor2006mass} & 0.078 \\
\cite{renedo2010new} & 0.118 \\
this work (MESA) & 0.049 \\

\hline
\end{tabular}
\label{4mod_fitqual}
\end{table}

\section{Conclusions}

We have described the implementation of a fitting statistic, known as the unbinned maximum likelihood, which is common in some fields but relatively unused in astronomy. This approach does not require that the data be binned in any way, or an assumption of Gaussian error bars, both of which are necessary for a $\chi^2$ fit. Given the same data and model, an unbinned approach will allow for tighter constraints on the same parameters than a binned fit will. In the case we have described, the error model is combined directly with the theoretical model to produce a continuous distribution for where points are expected to be found in data-space. The data can be kept unmodified from their raw state, and all assumptions about parameters and errors are built into the model itself. 

We used this approach to test white dwarf cooling models against the largest sample of white dwarfs ever collected in a single globular cluster (47 Tuc). Fitting in six-dimensional parameter space was performed with an MCMC sampler. We found constraints on the thickness of the Hydrogen layer and the amount of neutrino production in the white dwarf cores. 

The constraint on $q_\mathrm{H}$ is shown in Figure \ref{qhfit}. The data prefer thicker hydrogen layers $(q_\mathrm{H}=3.2\e{-5})$ and we can strongly rule out thin layers $(q_\mathrm{H}=10^{-6})$. The $95\%$ range of this parameter is $2.2\e{-5}$ to $8.9\e{-5}$. Our constraint on neutrino production is much better than for $q_\mathrm{H}$, with $0.83<A_{\nu}<1.22$ at $95\%$ confidence. In \cite{hansen2015constraining} the best-fitting hydrogen layer thickness was found to be $q_\mathrm{H}=4\e{-4}$. Our analysis strongly rules out that value, but we also find better fitting hydrogen layers at the thick end of the range usually considered by modellers \citep{bergeron2001photometric}. The $95\%$ confidence range on $f_s$ in \cite{hansen2015constraining} (which is equivalent to our parameter $A_{\nu}$) was found to be $0.6<f_s<1.7$, which agrees with what we have found. The differences in these values can be attributed both to the inclusion of additional data in our analysis, and to different statistical assumptions made in fitting the models.

\break

\bibliographystyle{apj}
\bibliography{unbin}

\end{document}